\documentclass[twocolumn,floatfix,superscriptaddress,amsmath,showpacs,showkeys,aps,prb]{revtex4-1}
\usepackage[utf8]{inputenc}
\usepackage[final]{graphicx}
\usepackage{caption}
\usepackage{subcaption}
\usepackage{epsfig}
\usepackage{bm}
\usepackage[normalem]{ulem}
\usepackage{times}

\newcommand{\beq}{\begin{equation}}
\newcommand{\eeq}{\end{equation}}
\newcommand{\beqa}{\begin{eqnarray}}
\newcommand{\eeqa}{\end{eqnarray}}

\newcommand{\veck}{\vec{k}}
\newcommand{\ddk}{\frac{d^2 k}{(2\pi)^2}}

\begin{document}

\title{Nematic Phase in two-dimensional frustrated systems with power law decaying interactions}

\author{Daniel G. Barci}
\affiliation{Departamento de F{\'\i}sica Te\'orica,
Universidade do Estado do Rio de Janeiro, Rua S\~ao Francisco Xavier 524, 20550-013  
Rio de Janeiro, Brazil}
\author{Leonardo Ribeiro}
\affiliation{Departamento de F{\'\i}sica Te\'orica,
Universidade do Estado do Rio de Janeiro, Rua S\~ao Francisco Xavier 524, 20550-013  
Rio de Janeiro, Brazil}

\author{Daniel A. Stariolo}
\affiliation{Departamento de F\'{\i}sica,
Universidade Federal do Rio Grande do Sul and\\
National Institute of Science and Technology for Complex Systems\\
CP 15051, 91501-970 Porto Alegre, RS, Brazil}

\date{June 4, 2013}

\begin{abstract}
We address the problem of orientational order in  frustrated interaction systems as a function of the relative range of the competing interactions.  
We study a spin  model Hamiltonian
with short range ferromagnetic interaction competing with an antiferromagnetic component that decays as a power law of the distance between 
spins, $1/r^\alpha$.
These systems may develop a nematic phase between the isotropic disordered and stripe phases.
We evaluate the nematic order parameter using a self-consistent  mean field calculation.
Our main result indicates that  the nematic phase exists, at mean-field level,  provided $0<\alpha<4$.   
We analytically compute the nematic critical temperature  and show that it  increases with the range of the interaction,  reaching  
its maximum near $\alpha\sim 0.5$.
We also compute a corse-grained effective Hamiltonian for long wave-length fluctuations.  For $0<\alpha<4$ the inverse susceptibility develops a set of continuous minima at wave vectors $|\vec k|=k_0(\alpha)$ which dictate the long distance physics of the system.  For $\alpha\to 4$, 
$k_0\to 0$, making the competition between interactions ineffective for greater values of $\alpha$.  
\end{abstract}

%64.60.De 	Statistical mechanics of model systems (Ising model, Potts model, field-theory models, Monte Carlo techniques, etc.) 
%75.70.Kw	Domain structure (including magnetic bubbles and vortices)
% 75.30.Kz	Magnetic phase boundaries (including classical and quantum magnetic transitions, metamagnetism, etc.)
%75.70.Ak 	Magnetic properties of monolayers and thin films
%75.70.Cn 	Magnetic properties of interfaces (multilayers, superlattices, heterostructures)
%75.70.-i 	Magnetic properties of thin films, surfaces, and interfaces

\maketitle

%%%%%%%%%%%%%%%
\section{Introduction}
%%%%%%%%%%%%%%%
Local  structures at different scales,  
that could break translational as well as rotational invariance, generally appear in  systems with competing interactions.  
There are a variety of examples,  ranging  from solid state systems, like ultrathin ferromagnetic films
~\cite{Pescia2010,PoVaPe2003,WoWuCh2005} and strongly correlated electron systems~\cite{KiFrEm1998,
FrKi1999,BaFrKiOg2002,Lawler2006}, to soft matter systems like Langmuir monolayers~\cite{SeMoGoWo1991},
block copolymers~\cite{VeHaAnTrHuChRe2005,RuBoBl2008},
colloids and soft core systems~\cite{MaPe2004,ImRe2006,GlGrKaKoSaZi2007}. Besides the
intrinsic interest raised by the complexity of the phase behavior, 
their detailed knowledge could be relevant to  understand basic phenomena such as high temperature superconductivity, and
also for technological applications like soft matter templates for nanoscale systems and
future spintronic devices.  

Competing interactions at different scales may give
rise to complex phases and patterns, like stripes, lamellae, bubbles and others~\cite{SeAn1995}.
In this  intricate phase  structure, the nematic phase\cite{BaSt2007}, an homogeneous however non-isotropic state,  plays an important role. 
It may appear as an intermediate phase between a fully disordered phase and a modulated phase. 
An interesting approach to study the phase transitions in systems with isotropic competing interactions was early present in Ref. \onlinecite{Br1975}. 
 Analyzing a generic Ginsburg-Landau model, whose main characteristic is the
presence of a minimum in the spectrum of Gaussian fluctuations at a non-zero wave vector,
 it was shown  that the model has a 
first order transition  to a modulated phase. In  recent works\cite{BaSt2007,BaSt2009}  we have shown that 
 pure symmetry considerations, in the context of the renormalization group,  lead to terms in the free energy that encode orientational order parameters. 
 We were able to find a nematic phase at temperatures above the critical temperature for  modulated phases.
We  have also found that   the 
isotropic-nematic phase transition in the two-dimensional continuum system with isotropic competing interactions is in  the
Kosterlitz-Thouless universality class~\cite{BaSt2009}.  More recently, we 
developed a method to compute the nematic order parameter in a  classical spin Hamiltonian system 
with competing interactions\cite{BaSt2011}. We have applied the method to the Ising frustrated dipole ferromagnet\cite{CaMiStTa2006} and to 
the short ranged biaxial next-nearest-neighbor Ising model (BNNNI), or $J1$-$J2$ model\cite{JiSeSa2012}. Interestingly, for the dipolar interaction model a nematic phase was found, 
while  in the $J1$-$J2$ model this phase does not  exist. This points to the relevance of  the interaction range to  develop a nematic phase. 

Then, a natural question we address in this article is about the  range of the  frustrating interaction necessary to produce an intermediate 
nematic phase in between the disordered isotropic and the striped phase.The necessity of long ranged interactions is frequently invoked,
 but the actual influence of the  relative range between the competing interactions  is still an open problem.
 To answer this question we study a spin  model Hamiltonian
with short range ferromagnetic interaction competing with an antiferromagnetic component that decays as a power law with the distance between 
spin sites , $1/r^\alpha$ (where $r$ is the distance between two spins, and $\alpha$ measure the range of the decay). 
For $\alpha=3$, this is  the known Ising frustrated dipolar model, while $\alpha=1$ is equivalent to 
the frustrated Coulomb model\cite{GrTaVi2000}.    We have evaluated the nematic order parameter using a self-consistent  mean field calculation. To do this, it is necessary 
to compute spin fluctuations, since the nematic order parameter is quadratic in the spin variable. 
Our main result indicates that  the nematic phase exists, at mean-field level,  provided $0<\alpha<4$.   In other words, if the frustrating interaction decays 
faster than $1/r^4$,  a pure orientational order is not possible. 
We have analytically computed the nematic critical temperature   for $0<\alpha<4$ and have shown that the temperature window between the stripe and the nematic phase
increases with the range of the interaction, reaching  its maximum near $\alpha\sim 0.5$.
We have also computed a corse-grained effective Hamiltonian for long wave-length fluctuations.  For $0<\alpha<4$ the inverse susceptibility develops a set of continuous minima at wave vectors $|\vec k|=k_0(\alpha)$ which rules the long distance physics of the system.  For $\alpha\to 4$, 
$k_0\to 0$, making the competition ineffective for greater values of $\alpha$.  Also, the stiffness of pattern formation is enhanced with the range of the competing interaction.
For ranges shorter than the dipolar interaction $ \alpha> 3$ it takes vey small values, signaling a possible instability of mean-field order. 

The paper is organized as follows: in section \ref{Model} we present our model and compute the long wave-length effective field theory for any value of $\alpha$.
In \S \ref{Nematic} we compute the nematic order parameter and the critical temperature.  
Finally we discuss our results and conclusions in \S \ref{conclusions}, leaving some technical details for appendix \ref{FT}.

%%%%%%%%%%%%%%%%%%%%%%%
\section{Model Hamiltonian and effective field theory}
\label{Model}
%%%%%%%%%%%%%%%%%%%%%%%%%%
We consider  a Hamiltonian written in terms of Ising variables $S_i=\pm 1$,  with competition 
between short-range ferromagnetic and long-ranged
antiferromagnetic interactions
\begin{equation}
H= -\frac{J}{2} \sum_{<i,j>} S_i S_j + \frac{g}{2} \sum_{(i,j)}
\frac{S_i S_j}{r^\alpha_{ij}} + \sum_{i}  B_i S_i .
 \label{Hamiltonian}
\end{equation}
\noindent The first sum runs over all pairs of nearest
neighbors spins in a two-dimensional lattice, while the second one runs over all pairs of spins
of the lattice; $r_{ij}$ is the distance, measured in lattice
units, between sites $i$ and $j$.  $J,g>0$ measure the ferromagnetic exchange and the long-ranged frustrating antiferromagnetic interaction respectively.  
The range of the latter  is controlled by the exponent  $\alpha$.   The last term is the energy associated with an external magnetic field $B_i$. 
We are interested in the regime of small frustration $g<J$, since this is the relevant regime in some applications such as ferromagnetic thin films 
with perpendicular anisotropy in which the
special  case of $\alpha=3$ (frustrated Ising-dipolar model) is usually considered to model the physical system.    

Next, we analyze the effective long-wavelength behavior of this model as a function of $\alpha$.
As usual, the thermodynamic properties in the canonical ensemble are defined in terms of the partition function $Z(B)={\rm Tr} \exp(-\beta H)$.
It is well known that long-ranged interactions with $\alpha<d$  ($d=2$ is the dimensionality in this work), may lead to inequivalence between the canonical and the mircrocanonical 
ensembles~\cite{Ruffo2001}. The essential reason for this behavior is that the energy necessary to produce a homogeneous ground state is
infinite in the thermodynamic limit. However,  in competitive models like the ones we are considering, the phase transitions are dominated by  the  modulation scale~\cite{Ruffo2005}. In these cases, the energy is additive and in principle the canonical ensemble can be safely used.

It is convenient to re-write the partition function in terms of   real variables on the lattice ($-\infty<\Phi_i<\infty $). This can be done 
by means of a Hubbard-Stratonovich transformation\cite{BiDoFiNe1995}.  Exactly summing up the Ising degrees of freedom $S_i$, we can re-write the partition function in terms of  an effective Hamiltonian written in the new variables\cite{BaSt2011}, 
\beqa
{\cal H}[\{\Phi\}]& =& 
\frac{1}{4}\sum_{ij}\Phi_iJ_{ij}\Phi_j - \frac{1}{2} \sum_i B_i \Phi_i -\nonumber \\
&-& \frac{1}{\beta}
               \sum_i \log{\cosh{\left(\beta\sum_j J_{ij}\Phi_j\right)}}.
\label{effective}
\eeqa
In this expression, $J_{ij}$ is the total interaction matrix. For nearest neighbors, it is essentially the constant $J$, while for all other components is $-g/r_{ij}^\alpha$.
The sums run over all pairs of sites in a two-dimensional square lattice, the inverse temperature  $\beta=1/ T$ and $B$ is an external magnetic field.  It is not difficult to find a relation between the  original discrete variables and the new continuous ones, by just differentiating   the partition function with respect  to the magnetic field in both representations. One immediately finds\cite{BiDoFiNe1995}:
\beqa
\langle S_i\rangle&=&\frac{1}{2} \langle \Phi_i \rangle \\
\langle S_iS_j \rangle &=&  -\frac{1}{2\beta} J_{ij}^{-1} +\frac{1}{4} \langle \Phi_i \Phi_j \rangle,
\label{spincorrel}
\eeqa

It is instructive to write the Hamiltonian of Eq. (\ref{effective}) in the disordered high temperature phase. In this regime, the entropic last term can be expanded in powers of 
$\beta\Phi$. Keeping the leading order term we find (for $B=0$), 
\beq
{\cal H}_d\sim 
\frac{1}{4}\sum_{ij}\Phi_i\left\{{\bf J} \cdot ({\bf I}-2 \beta {\bf J} ) \right\}_{ij}\Phi_j  
\eeq
where ${\bf J}$ is the matrix whose   components are $J_{ij}$, ${\bf I}$ is the identity matrix  and the ``dot'' indicates usual matrix product. 
In reciprocal space, the quadratic Hamiltonian in two dimensions takes the simpler form,  
\beq
{\cal H}_d  \sim 
\frac{1}{4}  \int_{BZ} \frac{d^2 k }{(2\pi)^2}\; \Phi(k)\  \chi^{-1}(k) \Phi(-k)
\eeq
in which $BZ$ indicates the first Brillouin zone and $\chi^{-1}(k)$ is the Fourier transform of  ${\bf J} \cdot ({\bf I}-2 \beta {\bf J} )$. 

For isotropic interactions and long-wave components ($ka<<1$, where $a$ is the lattice constant),   the inverse susceptibility $\chi^{-1}(k)$ in the  high-temperature phase  depends only on $k=|\vec k|$.   
For simple ferromagnets this function has a minimum at $k_0=0$,  and then $\chi^{-1}$ can be expanded in Taylor series 
to arrive at the corse-grained Hamiltonian 
\beq
{\cal H}_d  \sim 
  \int_{|k|<1/a} \frac{d^2 k }{(2\pi)^2}\;  \left\{r + \rho k^2\right\} |\Phi(k)|^2 , 
\label{ferromagnetic}
\eeq
 that correctly describes the ferromagnetic phase transition\cite{Amit1978}.   
In Eq. (\ref{ferromagnetic}),
\beqa 
r& =& \chi^{-1}(0) \\
\rho&=& \frac{1}{2!} \left. \frac{d^2\chi^{-1}}{dk^2} \right|_{k=0} \;.
\eeqa

 However, in frustrated systems  $\chi^{-1}(k)$ may develop  a minimum for a finite wave vector $\vec k=\vec k_0$. The effective long distance Hamiltonian will then be dominated by this scale. 
Expanding  the inverse susceptibility in  Taylor series, 
the coarse grained Hamiltonian is now given at leading order by 
\beq
{\cal H}_d  \sim 
  \int_{|k-k_0|<\Lambda} \frac{d^2 k }{(2\pi)^2}\;  \left\{r + \rho (k-k_0)^2\right\} |\Phi(k)|^2
\label{Brazovskii}
\eeq
where  $\Lambda$ is a cut-off and  
\beqa 
r& =& \chi^{-1}(k_0) \\
\rho&=& \frac{1}{2!} \left. \frac{d^2\chi^{-1}}{dk^2} \right|_{k=k_0} 
\eeqa
Therefore, the magnetic susceptibility $\chi(0)$ is always finite.  On the other hand, $\chi(k_0)\sim 1/r$. 
If $r\to 0$, the susceptibility at the wave vector $k_0$ diverges, signaling a tendency of the magnetization to form modulated structures 
with wave vector $k_0$. For high temperatures $r>0$, the correlation length  $\xi\sim 1/ \sqrt{r}$.  The system tends to form stripe domains 
with wavelength  $\lambda=2\pi/k_0$, whose area  is proportional to $1/r$.  In this way, the magnetic susceptibility $\chi(k_0)$ is essentially a measure of 
the area of each stripe domain. These domains are, in principle, uncorrelated at high temperatures.    
The Hamiltonian of Eq. (\ref{Brazovskii}) was  proposed  long time ago as an effective theory to study stripe phases\cite{Br1975}.  
More recently, a generalization of this model in the context of the renormalization group was studied \cite{BaSt2009}. 
It was shown that, in the continuum two-dimensional model, while the stripe long-ranged order cannot exist (at least for sufficiently short ranged 
interactions), 
a pure orientational nematic order may developed indicating an orientational order 
of domain walls of local stripe order. 

Let us now return to  our ``microscopic'' model (Eq. (\ref{Hamiltonian})) and   analyze the structure of the disordered susceptibility in terms of the range of the frustrating interaction. 
The interaction matrix  can be cast in the form $J_{ij}= J J^{\rm f}_{ij}-g J^{\rm a}_{ij}$, where  $J^{\rm f}_{ij}$ is a ferromagnetic short ranged  interaction  and $J^{\rm. a}_{ij}$ represents the antiferromagnetic  long-ranged interaction. The Fourier transform of the first nearest neighbors ferromagnetic part  
 in a square lattice is $J^{\rm f}(k)= \cos k_x+\cos k_y$, where for simplicity we consider the lattice spacing $a=1$.  For long wavelength with respect to the 
lattice spacing $k_x, k_y \ll 1$, this interaction turns out to be isotropic,  $J^{\rm f}(k)\sim 2- (1/2) (k_x^2+k_y^2)$. On the other hand, the antiferromagnetic part, in the same isotropic approximation,    takes the form\cite{Gelfand} , 
$J^{\rm a}(k)= 2^{1-\alpha} \Gamma(1-\alpha/2)/\Gamma(\alpha/2) k^{\alpha-2}$ (for $\alpha\ne \mbox{even}$). 
 In this way, the interaction in reciprocal space can be written as (see appendix (\ref{FT})):
\beq
\label{J(k)}
J(k)= 2J\left\{1- \frac{1}{4} k^2+\frac{1}{2}\left(\frac{g}{J}\right) \sigma(\alpha) \frac{k^{\alpha-2}}{\alpha-2} \right\}
\eeq 
where
$
\sigma(\alpha)= 2^{2-\alpha}\Gamma(2-\alpha/2)/\Gamma(\alpha/2)
$.
Notice that,  for $\alpha=3$, the antiferromagnetic part reduces to $g k$, the well known long-distance behavior of the dipolar model. 
Eq. (\ref{J(k)}) is not well defined for even values of $\alpha$. A careful treatment of these cases leads to logarithmic corrections 
that are considered in detailed in 
Appendix \ref{FT}. 
%%%%%%%%%%%%%
\begin{figure}
        \centering
        \begin{subfigure}[b]{0.3\textwidth}
                \centering
                \includegraphics[scale=0.6]{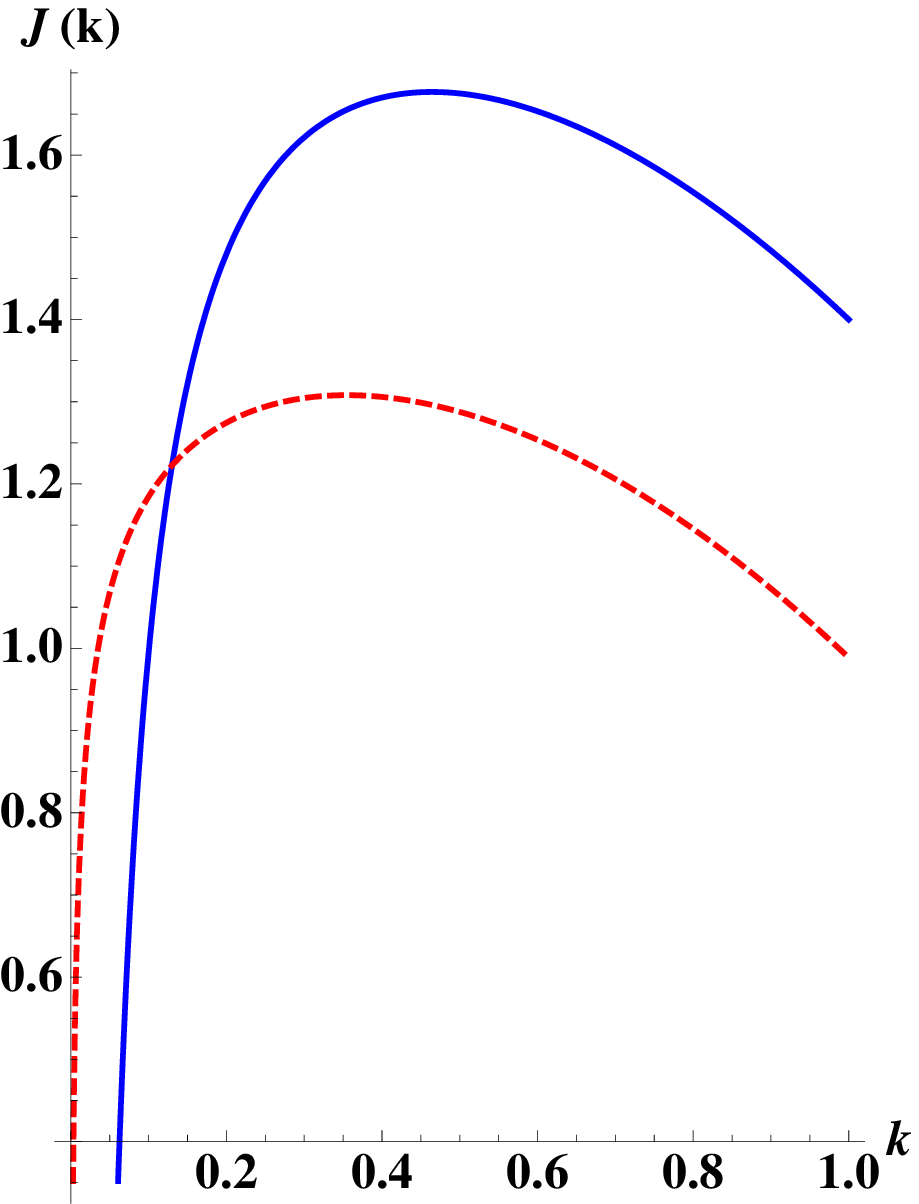}
                \caption{\label{figJl2}}
        \end{subfigure}
         \begin{subfigure}[b]{0.3\textwidth}
                \centering
                \includegraphics[scale=0.6]{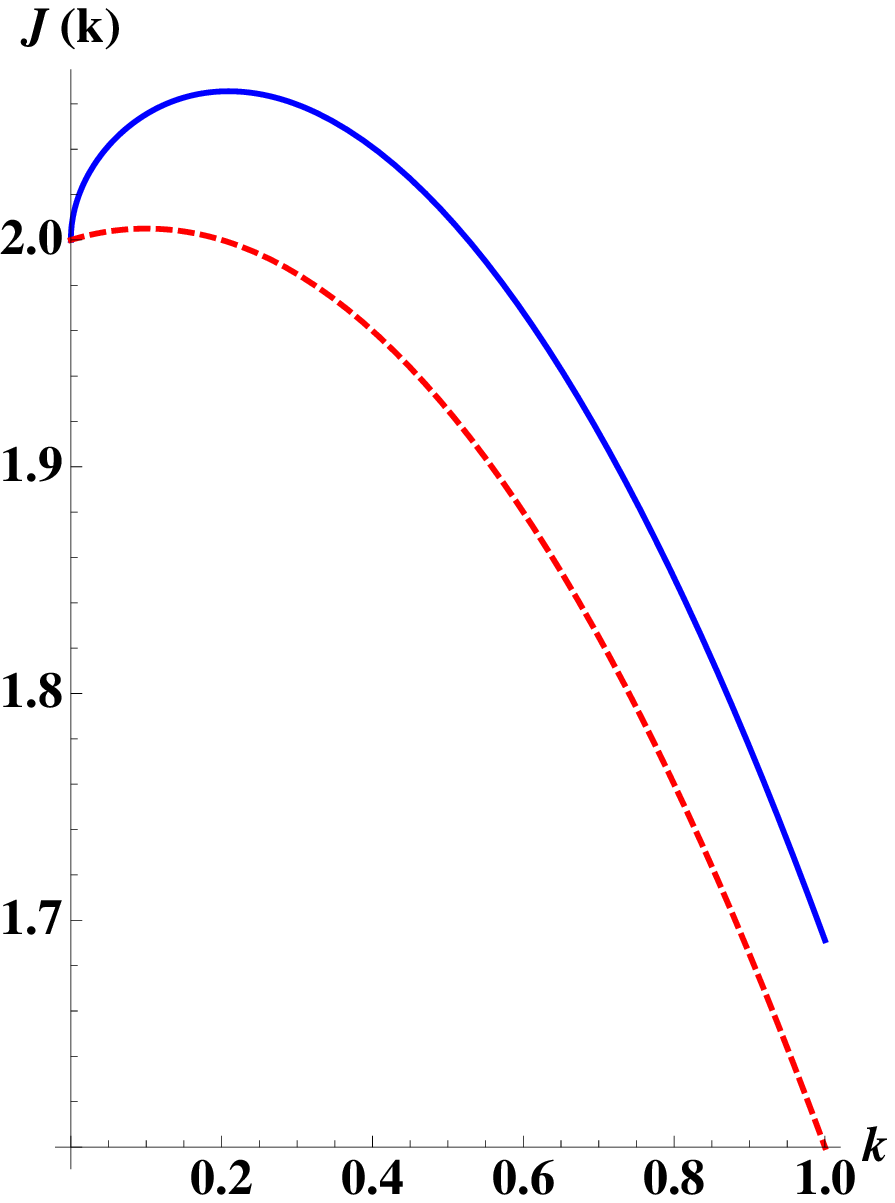}
                \caption{ \label{figJg2}}
        \end{subfigure}
\caption{$J(k)$ as a function of $k$ in units of $1/a$. We have fixed $J=1$ and $g=0.1$.
In Fig. (\ref{figJl2}),  $\alpha<2$. The bold line corresponds to $\alpha=1$, while the dashed line corresponds to $\alpha=1.8$. In Fig. (\ref{figJg2}), $\alpha>2$. 
The bold line corresponds to $\alpha=2.5$, while the dashed line corresponds to $\alpha=3$}
\label{figJ}
\end{figure}

\begin{figure}
\begin{center}
\includegraphics[scale=0.8]{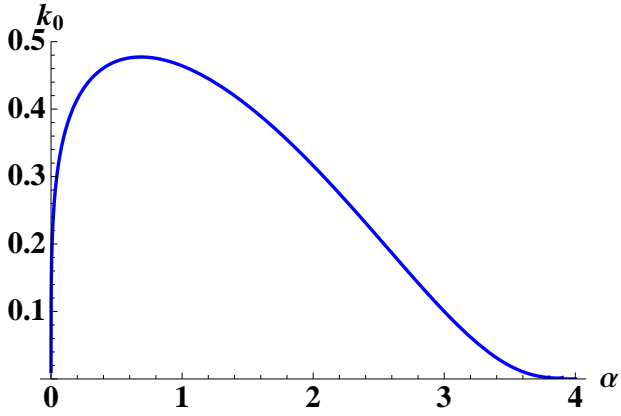}
\caption{ $k_0$ in units of the lattice spacing $a$, as a function of the parameter $\alpha$ for a fixed value of the competition parameter $g/J=0.1$}
\label{figk0}
\end{center}
\end{figure}

We see two special values, where the behavior of $J(k)$ changes qualitatively, $\alpha=2$ and $\alpha=4$.
From Eq. (\ref{J(k)}) we see that, for $\alpha> 4$, the antiferromagnetic term is sub-leading in the long distance limit ($k\ll1$), in such a way that $J(k)$ has a maximum at the value $k_0=0$. Then, the long-distance effective Hamiltonian is that of Eq. (\ref{ferromagnetic}) .
On the other hand, for  $\alpha<4$, we have two different situations. In the case  $\alpha<2$,   $\lim_{k\to 0}J(k)=-\infty$, while if $\alpha>2$, $J(0)=2J$. We illustrate these two cases in 
fig. (\ref{figJ}). In fig. (\ref{figJl2}) we depict two typical examples with $\alpha<2$, while in fig. (\ref{figJg2}) we show two cases with $\alpha>2$.  Despite the different behavior at the origin, we observe that for  $\alpha < 4$, $J(k)$ develops a maximum at a finite  scale $k_0\neq 0$ given by:
\beq
\label{k0}
k_0= \left[\sigma(\alpha)\left(\frac{g}{J} \right)\right]^{\frac{1}{4-\alpha}}
\eeq
In figure (\ref{figk0}) we depict the values of $k_0$, given by Eq. (\ref{k0}), as a function of $\alpha$ for a fixed value of $g/J=0.1$
We see that, in the regime of interest $g/J<1$,  a finite $k_0$ is developed all along   the interval $0<\alpha<4$.    After the dipolar value $\alpha > 3$,  $k_0$ rapidly decays to zero, and for  $\alpha>4$ it is no more possible to have $k_0\neq 0$. At the particular point $\alpha=4$   Eq. (\ref{J(k)}) is not well defined, since $\sigma(\alpha)$ has a pole. We have studied in detail this special case in  Appendix \ref{FT} and have shown that in this case  $k_0$ is exponentially small $k_0\sim \exp(-J/g)$. Then it can be considered zero for any practical purpose.  Also, the limit of $\alpha\to 2$ is not well defined in Eq. (\ref{J(k)}). However, upon differentiation, the value of $k_0$ is perfectly well defined. This is also discussed  in  Appendix \ref{FT}.    
Then, for $\alpha<4$, we can expand Eq. (\ref{J(k)}) in powers of $k-k_0$ obtaining:  
\beq
\label{J}
J(k)= J\left[ r_0 -\rho_0 \left(k-k_0\right)^2 \right]+ O((k-k_0)^3)
\eeq 
where  $r_0= J(k_0)/J\sim 2+ O(k_0^2) $ and    $\rho_0=(4-\alpha)/2$.
The inverse susceptibility in the same approximation reads :
\beqa
\label{chi1}
 \frac{1}{J}\;\chi^{-1}(k)& =& J(k)\left(1-2\beta J(k)\right) \\ 
&=& r+ \rho (k-k_0)^2+  \ldots \; .
\label{chi2}
\eeqa
Using Eq. (\ref{J})  we obtain for the coefficients:
\beqa
\label{r}
r&=& r_0 (1- 2r_0\beta J)  \\
\label{rho}
\rho&=&\rho_0 (4r_0\beta J-1)
\eeqa
We see that, for a high-temperature regime, $\rho$ could be negative, indicating an instability of  the theory. 
This is a very well known limitation of the method, that has its origin in the use of the Hubbard-Stratonvich transformation for 
kernels which are not positive definite\cite{Amit1978}. However, we are interest in the temperature regime in which the systems 
has a tendency to form patterns, {\em i. e. } $0<r\ll 1$ in which $\rho\sim \rho_0$. 
 In this regime the effective Hamiltonian Eq. (\ref{Brazovskii}), with the parameters $k_0, r, \rho$  given by Eqs. 
 (\ref{k0}), (\ref{r}) and (\ref{rho}) respectively, is perfectly well defined.  
 
\begin{figure}
\includegraphics[scale=0.8]{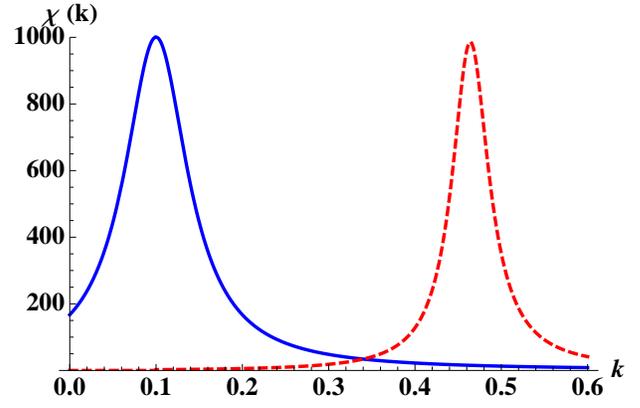}
\caption{ Magnetic susceptibility $\chi(k)$ as given by Eq. (\ref{chi1})  for two typical values of $\alpha$.  The bold line corresponds to frustrated dipolar Ising model, $\alpha=3$ with $\beta J=1/4.012$  while the dashed line 
is the susceptibility of the frustrated Coulomb Ising model,  $\alpha=1$,with  $\beta J=1/3.3557$. In both cases  we fixed $J=1$  and $g=0.1$. }
\label{figchi}
\end{figure}

 In fact, the expansion in $k-k_0$, given by Eq. (\ref{chi2}) is an excellent  approximation of Eq. (\ref{chi1}) near the the temperature $\beta_{st}\sim1/2r_0 J$,  where the 
 susceptibility diverges at $k=k_0$,   signaling the tendency to form striped patterns. In fig. (\ref{figchi})  we show two examples of the susceptibility computed from Eq. (\ref{chi1}).
 The bold line corresponds to frustrated dipolar Ising model, $\alpha=3$ at  $\beta J=1/ 4.012$,   while the dashed line 
represents  the susceptibility of the Coulomb model,  $\alpha=1$ at   $\beta J=1/3.3557$. We see that both curves are sharply picked at the corresponding value of $k_0$, given 
by fig. (\ref{figk0}),  while the widths of the peaks are proportional to $1/\rho$.  Moreover, the width of the curve $\alpha=3$ is clearly larger than the one with $\alpha=1$. This happens because the stiffness  $\rho$ grows with the range of the competing interaction,({\em i.e.}, with decreasing $\alpha$ ). In fact, from  Eq. (\ref{rho}),  we can observe  that $\rho\sim\rho_0\sim (4-\alpha)$.  Finally, the small temperature difference between both examples are due to the quadratic corrections of $r_0\sim 2+ O(k_0^2)$ that makes the stripe critical temperature  $\beta_{\rm st}$,  $\alpha$-dependent.
%%%%
%%%%

%%%%%%%%%%%%%%%%%%%%%%%%%%%%%%%%%%%%%%%%%%%%%%%%%%%%%%%%%%%
\section{Mean field theory for the nematic order parameter}
\label{Nematic}
%%%%%%%%%%%%%%%%%%%%%%%%%%%%%%%%%%%%%%%%%%%%%%%%%%%%%%%%%%%
The  orientational (nematic) order parameter is defined as\cite{BaSt2011}:
\beq
Q=\sum_{ij} \langle S_i S_{i+\hat x}-S_i S_{i+\hat y}\rangle ,
\label{Q1}
\eeq
where $\hat x$ and $\hat y$ are unit vectors along the $x$ and $y$ axes of the square lattice.
With this definition, the usual tensor nematic order parameter has only one component. 
If $Q$  is positive, the director points along
the $x$ direction while if it is negative, the director mainly points in the $y$ direction. 
These are the only two possible directions of the director. For this reason, if $Q\neq 0$, the resulting anisotropic phase is called Ising-Nematics, since it breaks the rotational point group of the lattice and it is invariant under rotations by $\pi$. 
On the other hand, in the continuum  limit, the nematic order parameter acquire a phase,  $Q=|Q| e^{i2\theta}$, since the direction in the plane is now arbitrary. In this case, 
if $Q\ne 0$, the director can point in any direction labeled by $\theta$. The factor of $2$ in the exponent guarantees the nematic symmetry $\theta\to\theta+\pi$.
Along the paper, for brevity,  we generally use the term ``nematic'' to refer to any of these phases, however, whenever we deal with a square lattice  model, ``Ising-Nematic''  should be understood.  

Eq. (\ref{Q1}) can be written as 
\beq
 Q = \frac{1}{2}\sum_{ij} K_{ij} \langle S_iS_j\rangle= \frac{1}{8}\sum_{ij} K_{ij} \langle \Phi_i\Phi_j \rangle .
\label{globalnematicparameter}
\eeq
where
\beq
K_{ij}=\left\{ 
\begin{array}{lcl}
+1  & \mbox{if}  & j=i\pm \hat x \\
-1  & \mbox{if}  & j=i\pm \hat y
\end{array}
\right.
\eeq
In the last equality we used Eq. (\ref{spincorrel}) and the fact that $Tr ({\bf K}\cdot {\bf J}^{-1})=0$ by symmetry.

Therefore, to compute $Q$ we need to evaluate the correlation function $\langle \Phi_i\Phi_j \rangle$ in the ordered ($Q\neq 0$) phase. 
We referred the reader to  reference \onlinecite{BaSt2011} for  a detailed mathematical formalism to compute this quantity, based on the Hamiltonian 
Eq. (\ref{effective}).  Here, we sketch the main physical concepts behind this formalism.
In the nematic ordered phase (if it exists), the system is homogeneous, however having a global anisotropy given by $Q$.  The order parameter acts as 
a nematic mean field, in such a way that  individual magnetic moments couple with  $Q$,  contributing to the mean field Hamiltonian with an energy 
$JQ \Phi_iK_{ij} \Phi_j$. This the simplest way that $Q$ can be coupled to the magnetic moments  satisfying rotational invariance. On the other hand,  it is the leading order term 
in an homogeneous  phase $\langle\Phi_i\rangle=0$.  Therefore, the  mean field Hamiltonian in the nematic  phase has the form, 
\beq
{\cal H}_{\rm nematic}\sim 
\frac{1}{4}\sum_{ij}\Phi_i\left\{{\bf J} \cdot ({\bf I}-2 \beta {\bf J} )-2J Q {\bf K} \right\}_{ij}\Phi_j  
\eeq
The (anisotropic) correlation matrix is
\beq
\langle \Phi_i\Phi_j \rangle\sim\left( {\bf J} \cdot ({\bf I}-2 \beta {\bf J} )-2J Q {\bf K} \right)_{ij}^{-1}   \; .
\eeq
Substituting this expression into the definition of the order parameter (Eq. (\ref{globalnematicparameter})) we find a self-consistent equation 
for $Q$ given by\cite{BaSt2011},  
\beq
Q= \frac{1}{16\beta}\  {\rm Tr}\left\{ \frac{{\bf K}}{{\bf J}-2\beta {\bf J}^2-2J Q{\bf K}}\right\}.
\label{Qmf}
\eeq
This  equation is the analog of the Curie-Weiss approximation
for the magnetization in the Ising model. 
If this equation has a non-trivial solution 
$Q\neq 0$, then the system exhibits an anisotropic but otherwise homogeneous phase with nematic symmetry. 
The presence or not of this phase depends on the detailed structure of the competing interactions, 
coded in the explicit form of the matrix ${\bf J}$.
We can look for a critical point by expanding 
the r.h.s. of (\ref{Qmf}) in  powers of $Q$, to obtain
\beq
16\beta Q \approx  2J Q  Tr \left( {\boldsymbol \chi}{\bf K}  \right)^2
         + 8J^3 \; Q^3 Tr \left( {\boldsymbol \chi}{\bf K} \right)^4.
\eeq
where ${\boldsymbol \chi}$ is the magnetic susceptibility matrix in the disordered isotropic phase. 
$Q=0$ is always a solution of the self-consistent  equation. If $Q\neq 0$ then for $Q \ll1$:
\beq
Q \approx \frac{1}{2} \left[ \frac{8\beta J - 
Tr \left( {\boldsymbol \chi}{\bf K} \right)^2}{Tr \left( {\boldsymbol \chi}{\bf K} \right)^4} \right]^{1/2}
\label{Qaprox}
\eeq
This result implies a continuous, second order {\em isotropic-nematic} transition, at a critical 
temperature given by:
\beq
\beta_c=\frac{1}{8}\  Tr \left({\boldsymbol \chi}(\beta_c){\bf K}\right)^2.
\label{Tcritica}
\eeq
In reciprocal space eq. (\ref{Tcritica}) reads:
\beq
\beta_c=\frac{1}{8}\  \int_{\rm BZ} \ddk \left[ \chi_c(\veck) K(\veck)      \right]^2.
\label{Tcont}
\eeq
where,  
$K(\veck) =2(\cos{k_x }-\cos{k_y })$. In ref. \onlinecite{BaSt2011} we have numerically 
solved this equation in the case of the Ising-frustrated-dipolar model, i.e. for $\alpha=3$, where the conditions for the existence 
of the nematic transition in terms of the microscopic parameters $J$ and $g$ was shown.   
Here, we show an approximate analytic solution for any value of $\alpha$.  To do this, we write a long-distance (continuous) approximation 
of Eq. (\ref{Tcont}) and we focus in the regime where the temperature is very near the instability towards stripe formation $r<<1$.   
At long distances,  $K(\vec k) \sim  k_x^2-k_y^2 = k^2 \cos(2\theta)$.
Using the results for the magnetic susceptibility found in Eq. (\ref{chi2}) we find for the nematic critical temperature, 
\beq
J\beta_c=\frac{1}{8}\  \int \ddk \left[ \frac{ k^2 \cos(2\theta)} {r_c+\rho_c(k-k_0)^2} \right]^2.
\label{Tld}
\eeq
where $r_c$ and $\rho_c$ are given by Eqs. (\ref{r}) and (\ref{rho}) at $\beta=\beta_c$.
Performing the angular integration we finally find,  in dimensionless quantities, the following self-consistent equation for the critical temperature
\beq
8\pi J\beta_c=  \int_0^1 dk    \frac{ k^5} {\left[r_c+\rho_c(k-k_0)^2 \right]^2}  \;\; .
\label{betac}
\eeq
Note that if $k_0=0$, the integral is completely regular, and almost temperature independent, dominated by the ultraviolet cut-off. However, 
for finite $k_0 \neq 0$, the integrand for $k\to k_0$  diverges as $r_c\to 0$. 
Then,  the integral can be approximated at leading order in  $r_c\ll 1$, and  $\rho_c\sim \rho_0$,  
\beq
\int_0^1 dk    \frac{ k^5} {\left[r_c+\rho_c(k-k_0)^2 \right]^2} \sim \frac{8k_0^5}{3\rho_0^{1/2}} \frac{1}{r_c^{3/2}} + O (r_c^{-1/2}).
\label{integral}
\eeq
Substituting Eq. (\ref{integral}) into Eq. (\ref{betac}) and using Eqs. (\ref{r}) and (\ref{rho}) we find
\beq
 \beta_c= \frac{1}{2 r_0 J}\left(1-\delta\right) 
\eeq
where we have defined a small quantity $0<\delta\ll 1$
\beq
\delta=\frac{\beta_{\rm st}-\beta_c}{\beta_{\rm st}}=\left[ \frac{2}{9\pi^2} \frac{k_0^{10}}{\rho_0}\right]^{1/3}.
\label{delta}
\eeq
Thus, for small frustration $g/J<1$ and for long ranged interactions $\alpha<4$,  the model of Eq. (\ref{Hamiltonian}) presents a   homogeneous but anisotropic phase 
with nematic symmetry.   Note that, at the  isotropic/nematic critical temperature $r_c=2\delta\ll 1$ and  the magnetic susceptibility  at wave-vector  $k_0$,  $\chi(k_0)\sim 1/\delta$ is 
finite, showing that the system is homogeneous.   
For interaction ranges longer than the dipolar model ($\alpha < 3$) the temperature window for the nematic phase grows. Conversely, 
for shorter range interactions ($\alpha>3$)  the nematic critical temperature decreases.  
Furthermore,   as  $\alpha\to 4$ not only the critical temperature decreases, but also the modulation wavelength grows rapidly, $\lambda=2\pi/k_0\to \infty$. 
In practice, there is an upper limit  in the value of $\alpha\lesssim 4 $, given by the  finite size of the sample. In fact, for  $k_0(\alpha) L\sim 1$ (where $L$ is the linear dimension of the sample), the stripe wavelength $\lambda\sim L$.  Near these values  $\chi(k_0)\sim\chi(0)$ and the size of the domains are of the same order than the system size.
Thus, the competition turns out to be ineffective and the system behaves, at long distances,  as a uniform ferromagnet.

%%%%%%%%%%%%%%%%%%%%%%%%%%%%%%%%%%%%%
\section{Summary and conclusions}
\label{conclusions}

We have addressed the role of the  relative interaction ranges on  the nematic phase in competing interaction  models at different scales.
We have studied a two dimensional Ising model on a square lattice, with short ranged ferromagnetic interaction and long-ranged 
antiferromagnetic one, whose range decays as a power law controlled by a parameter $\alpha$.

We have mapped the microscopic model into a corse-grained field theory that describes the long-distance behavior. 
We showed that, for small frustration $g/J<1$ and long-ranged interactions $\alpha<4$, the system develops a scale $k_0$ which dominates the 
low-energy physics.  In this regime the effective field theory is formally equivalent to a Brazovskii model in two dimensions.  On the other hand, for shorter ranged interactions 
 $\alpha>4$, $k_0=0$ and the system behaves as a usual ferromagnet.  

We focused on the isotropic-nematic transition, looking for an homogeneous  phase with anisotropic correlations. 
To compute the nematic order parameter  we used a self-consistent approach, previously applied\cite{BaSt2011} to the frustrated-dipolar-Ising interaction.
We have shown that there is a temperature window above the stripe instability in which the nematic phase can be  developed for ranges of $\alpha<4$. This window grows 
with the range of the frustrating interaction up to very long ranges $\alpha<1$.  On the other hand, for ranges shorter than the dipolar interaction $\alpha>3$ the critical temperature 
gets smaller and the stripe  wavelength  grows rapidly.  For $\alpha\sim 4$ the domain size is very large and in the limit of $k_0 L\sim 1$ it is  of the  the same order of the system size. In this regime the competition turns out to be irrelevant.  
Although for $ 3<\alpha<4$  mean field theory predicts a nematic phase,   fluctuations may destroy nematic order since the ``stiffness''  $\rho_0$ 
 is much weaker than in the region $\alpha\le 3$.  For $\alpha>4$ there is  neither stripe nor nematic solution even at mean field. 
 A word of caution is in order here.  The threshold $\alpha=4$ for the existence of $k_0\neq 0$ is valid in the asymptotic small frustration limit $g/J\ll 1$. If $g\sim J$, the system could develop  a finite scale $\vec k_0\neq 0$  very near the edge of the Brillouin zone. In this case, the isotropic approximation of the antiferromagnetic interaction is no longer valid.  For strong frustration, it is necessary to take into account the short-ranged part of the competing interaction.    
  This type of interactions, like for instance in the ANNNI or BNNNI models are 
 necessarily anisotropic since at least have the discrete symmetry of the lattice. These systems, although with short range interactions, develop a finite scale $k_0$ which leads to the appearance of a stripe phase. However, the susceptibility $\chi(\vec k)$ has a small number of isolated maxima\cite{BaSt2011}, differently from our model in which we have an infinitely degenerate set of maxima. This 
 difference is at the heart of the existence of the nematic phase as we have studied in the present work. 
  
Interesting enough, our result for the modulation period $k_0(\alpha)$ ( Eq. (\ref{k0}) and Fig. (\ref{figk0})) is in complete agreement with  very general scaling properties studied  in Ref. \onlinecite{Portman2010}.  The scaling behavior of the modulation length with $g/J$ resides on the homogeneity of  the long-ranged interaction in Fourier space  (Eq. (\ref{fa})).  Indeed, it was shown\cite{Portman2010} that, for $T=0$, the threshold exponent for the occurrence of modulated ground states is 
$\alpha=3$ (for $d=2$), while at high temperatures we have shown a very similar behavior  at $\alpha=4$. Thus, the scaling behavior of $k_0$ as a function of $g/J$  is strongly temperature dependent. This could imply that, for models in the range $3\leq\alpha<4$, it could exist a critical temperature below which a disorder phase reappears from a high temperature modulated phase. In fact, this reentrant behavior has been experimentally observed, see for instance Ref. \onlinecite{Pescia2010}.  Of course, in the context of the present work, we are not able to confirm this conjecture,  since our calculation is restricted to high temperatures very near the critical point.

Summarizing, this paper  is a contribution to understand a sector of a complex phase diagram in general models with competing interactions at different scales. 
The results presented have to be considered as a qualitative guide to more precise calculations. The main difficulty to compute 
quantitative relevant results is that, in general,  orientational order parameters in these kind of systems are quadratic 
functions of fundamental degrees of freedom. Even at mean-field level the computation of the order parameter implies the evaluation of fluctuations 
in the original variables.

%%%%%%%%%%%
\acknowledgments
The Brazilian agencies {\it Conselho Nacional de Pesquisas Cient\'\i ficas e T\'ecnicas} (CNPq), and {\it Funda\c c\~ao de Amparo \`a Pesquisa do Estado do Rio de Janeiro } (FAPERJ) are 
acknowledged for financial  support.

%%%%%%%%%%%%%%%%%
\appendix
\section{Fourier transform of  long-ranged interaction kernel}
\label{FT}
%%%%%%%%%%%%%%%%
The antiferromagnetic contribution to quadratic part of the effective Hamiltonian can be written as 
\beq
H^{a}_2=\frac{g}{2} \sum_{ij} \Phi_i \frac{1}{|\vec r_i-\vec r_j|^\alpha} \Phi_j\; .
\eeq
In reciprocal space, it reads 
\beq
H^{a}_2=\frac{g}{2} \int_{BZ} \frac{d^2k}{(2\pi)^2}  f^a(\vec k) |\Phi(k)|^2 \; .
\eeq
Assuming that the most relevant contribution to the phase transition comes  from length-waves much longer than the lattice spacing, 
we can consider the continuum limit, in which 
\beq
f^a(\vec k)=\int d^2 r  \frac{1}{|\vec r|^\alpha}  \; e^{i\vec k\cdot \vec r}
\eeq
This expression  coincides with  the exact form of $f^a(\vec k)$ for small values of $|\vec k| a << 1$, where $a$ is the lattice spacing.

The Fourier transform of $1/|\vec r|^\alpha$ in $d$ dimensions and for $\alpha$ not even is given by\cite{Gelfand},  
\beq
{\cal F}(1/r^\alpha)= 2^{d-\alpha}\frac{\Gamma(\frac{d-\alpha}{2})}{\Gamma(\frac{\alpha}{2})}  k^{\alpha-d}
\eeq
where the coefficient is written in terms of usual Gamma functions. 
Then, in two dimensions, and for $\alpha\neq 2, 4, \ldots$.
\beq
f^a(k)= 2^{2-\alpha}\frac{\Gamma(\frac{2-\alpha}{2})}{\Gamma(\frac{\alpha}{2})}  k^{\alpha-2}
\label{fa}
\eeq
Multiplying a dividing by   $(2-\alpha)/2$ and using the Gamma function property $z \Gamma(z)=\Gamma(z+1)$ we finally arrive at 
\beq
f^a(k)=-\sigma(\alpha) \frac{ k^{\alpha-2}}{\alpha-2}
\label{fa1}
\eeq
where $\sigma(\alpha)= 2^{2-\alpha}\Gamma(\frac{4-\alpha}{2})/\Gamma(\frac{\alpha}{2})$.
Eq. (\ref{fa1}) was used to build up  the long-wavelength  interaction in reciprocal space, Eq. (\ref{J(k)}).
 
As we have seen, for $\alpha > 4$, the exponent of $k$ is greater  than $2$, then at small $k$ the antiferromagnetic component 
is irrelevant with respect to the ferromagnetic one. 
Conversely,  for $\alpha\le 4$ this term essentially changes the behavior of $J(k)$. However, 
note that the Gamma function has poles at zero and negative integer values. Then, Eq. (\ref{fa}) is not well defined for $\alpha=2,4$.

The Fourier transform of negative integer exponents in two dimensions reads\cite{Gelfand} 
\beq
{\cal F}\left(\frac{2\pi }{r^{2 m + 2}}\right)= c_{-1} ^{(2+2 m)} k^{2 m }\ln k +c_{0} ^{(2+2 m)} k^{2 m}
\eeq
where $m$ is a positive integer and  the coefficients $c_0$ and $c_{-1}$ arise form the Laurent expansion of the the Gamma functions in Eq. (\ref{fa})
\beqa
2^{2-\alpha}\pi\frac{\Gamma(\frac{2-\alpha}{2})}{\Gamma(\alpha/2)}&=& \frac{c_{-1}^{(2+2m)}}{2-\alpha+2m}+ c_0^{(2+2m)}+ \nonumber \\
&+&c_1^{(2+2m)} (2-\alpha+2m)\ldots
\label{Laurent}
\eeqa
The case $\alpha=2$ corresponds to $m=0$ in the preceding equation giving:
\beq
\lim_{\alpha\to 2 } f^a(k)= -\ln k + 1/20
\eeq
Using this expression to compute $J(k)$, we find $k_0=\sqrt{g/J}$ that  coincides with Eq. (\ref{k0}) for $\alpha=2$.
Then, the curve $k_0(\alpha)$ depicted in figure (\ref{figk0}) is continuous at $\alpha=2$.

The other potentially problematic point is $\alpha=4$. This corresponds to the value $m=1$ in Eq. (\ref{Laurent}). In this case, 
\beq
\lim_{\alpha\to 4 } f^a(k)= -k^2 (1-\ln k )
\eeq
Using this expression to built up $J(k)$,   we find an exponentially small value of  
$k_0\sim \exp(-\frac{J}{2g})$. Also, the value of the stiffness in this case is also very small $\rho_0\sim g/J$. Then, for all 
practical proposes,  this limiting case can be safely ignored and $k_0(\alpha)$ is correctly represented in fig. (\ref{figk0}) 
for the entire range of $\alpha$.

%\bibliographystyle{apsrev4-1}
%\bibliography{ultrathin,nematics}

\begin{thebibliography}{27}%
\makeatletter
\providecommand \@ifxundefined [1]{%
 \@ifx{#1\undefined}
}%
\providecommand \@ifnum [1]{%
 \ifnum #1\expandafter \@firstoftwo
 \else \expandafter \@secondoftwo
 \fi
}%
\providecommand \@ifx [1]{%
 \ifx #1\expandafter \@firstoftwo
 \else \expandafter \@secondoftwo
 \fi
}%
\providecommand \natexlab [1]{#1}%
\providecommand \enquote  [1]{``#1''}%
\providecommand \bibnamefont  [1]{#1}%
\providecommand \bibfnamefont [1]{#1}%
\providecommand \citenamefont [1]{#1}%
\providecommand \href@noop [0]{\@secondoftwo}%
\providecommand \href [0]{\begingroup \@sanitize@url \@href}%
\providecommand \@href[1]{\@@startlink{#1}\@@href}%
\providecommand \@@href[1]{\endgroup#1\@@endlink}%
\providecommand \@sanitize@url [0]{\catcode `\\12\catcode `\$12\catcode
  `\&12\catcode `\#12\catcode `\^12\catcode `\_12\catcode `\%12\relax}%
\providecommand \@@startlink[1]{}%
\providecommand \@@endlink[0]{}%
\providecommand \url  [0]{\begingroup\@sanitize@url \@url }%
\providecommand \@url [1]{\endgroup\@href {#1}{\urlprefix }}%
\providecommand \urlprefix  [0]{URL }%
\providecommand \Eprint [0]{\href }%
\providecommand \doibase [0]{http://dx.doi.org/}%
\providecommand \selectlanguage [0]{\@gobble}%
\providecommand \bibinfo  [0]{\@secondoftwo}%
\providecommand \bibfield  [0]{\@secondoftwo}%
\providecommand \translation [1]{[#1]}%
\providecommand \BibitemOpen [0]{}%
\providecommand \bibitemStop [0]{}%
\providecommand \bibitemNoStop [0]{.\EOS\space}%
\providecommand \EOS [0]{\spacefactor3000\relax}%
\providecommand \BibitemShut  [1]{\csname bibitem#1\endcsname}%
\let\auto@bib@innerbib\@empty
%</preamble>
\bibitem [{\citenamefont {Saratz}\ \emph {et~al.}(2010)\citenamefont {Saratz},
  \citenamefont {Lichtenberger}, \citenamefont {Portmann}, \citenamefont
  {Ramsperger}, \citenamefont {Vindigni},\ and\ \citenamefont
  {Pescia}}]{Pescia2010}%
  \BibitemOpen
  \bibfield  {author} {\bibinfo {author} {\bibfnamefont {N.}~\bibnamefont
  {Saratz}}, \bibinfo {author} {\bibfnamefont {A.}~\bibnamefont
  {Lichtenberger}}, \bibinfo {author} {\bibfnamefont {O.}~\bibnamefont
  {Portmann}}, \bibinfo {author} {\bibfnamefont {U.}~\bibnamefont
  {Ramsperger}}, \bibinfo {author} {\bibfnamefont {A.}~\bibnamefont
  {Vindigni}}, \ and\ \bibinfo {author} {\bibfnamefont {D.}~\bibnamefont
  {Pescia}},\ }\href {\doibase 10.1103/PhysRevLett.104.077203} {\bibfield
  {journal} {\bibinfo  {journal} {Phys. Rev. Lett.}\ }\textbf {\bibinfo
  {volume} {104}},\ \bibinfo {pages} {077203} (\bibinfo {year}
  {2010})}\BibitemShut {NoStop}%
\bibitem [{\citenamefont {Portmann}\ \emph {et~al.}(2003)\citenamefont
  {Portmann}, \citenamefont {Vaterlaus},\ and\ \citenamefont
  {Pescia}}]{PoVaPe2003}%
  \BibitemOpen
  \bibfield  {author} {\bibinfo {author} {\bibfnamefont {O.}~\bibnamefont
  {Portmann}}, \bibinfo {author} {\bibfnamefont {A.}~\bibnamefont {Vaterlaus}},
  \ and\ \bibinfo {author} {\bibfnamefont {D.}~\bibnamefont {Pescia}},\
  }\href@noop {} {\bibfield  {journal} {\bibinfo  {journal} {Nature}\ }\textbf
  {\bibinfo {volume} {422}},\ \bibinfo {pages} {701} (\bibinfo {year}
  {2003})}\BibitemShut {NoStop}%
\bibitem [{\citenamefont {Won}\ \emph {et~al.}(2005)\citenamefont {Won},
  \citenamefont {Wu}, \citenamefont {Choi}, \citenamefont {Kim}, \citenamefont
  {Scholl}, \citenamefont {Doran}, \citenamefont {Owens}, \citenamefont {Wu},
  \citenamefont {Jin},\ and\ \citenamefont {Qiu}}]{WoWuCh2005}%
  \BibitemOpen
  \bibfield  {author} {\bibinfo {author} {\bibfnamefont {C.}~\bibnamefont
  {Won}}, \bibinfo {author} {\bibfnamefont {Y.~Z.}\ \bibnamefont {Wu}},
  \bibinfo {author} {\bibfnamefont {J.}~\bibnamefont {Choi}}, \bibinfo {author}
  {\bibfnamefont {W.}~\bibnamefont {Kim}}, \bibinfo {author} {\bibfnamefont
  {A.}~\bibnamefont {Scholl}}, \bibinfo {author} {\bibfnamefont
  {A.}~\bibnamefont {Doran}}, \bibinfo {author} {\bibfnamefont
  {T.}~\bibnamefont {Owens}}, \bibinfo {author} {\bibfnamefont
  {J.}~\bibnamefont {Wu}}, \bibinfo {author} {\bibfnamefont {X.~F.}\
  \bibnamefont {Jin}}, \ and\ \bibinfo {author} {\bibfnamefont {Z.~Q.}\
  \bibnamefont {Qiu}},\ }\href@noop {} {\bibfield  {journal} {\bibinfo
  {journal} {Phys. Rev. B}\ }\textbf {\bibinfo {volume} {71}},\ \bibinfo
  {pages} {224429} (\bibinfo {year} {2005})}\BibitemShut {NoStop}%
\bibitem [{\citenamefont {Kivelson}\ \emph {et~al.}(1998)\citenamefont
  {Kivelson}, \citenamefont {Fradkin},\ and\ \citenamefont
  {Emery}}]{KiFrEm1998}%
  \BibitemOpen
  \bibfield  {author} {\bibinfo {author} {\bibfnamefont {S.~A.}\ \bibnamefont
  {Kivelson}}, \bibinfo {author} {\bibfnamefont {E.}~\bibnamefont {Fradkin}}, \
  and\ \bibinfo {author} {\bibfnamefont {V.~J.}\ \bibnamefont {Emery}},\
  }\href@noop {} {\bibfield  {journal} {\bibinfo  {journal} {Nature}\ }\textbf
  {\bibinfo {volume} {393}},\ \bibinfo {pages} {550} (\bibinfo {year}
  {1998})}\BibitemShut {NoStop}%
\bibitem [{\citenamefont {Fradkin}\ and\ \citenamefont
  {Kivelson}(1999)}]{FrKi1999}%
  \BibitemOpen
  \bibfield  {author} {\bibinfo {author} {\bibfnamefont {E.}~\bibnamefont
  {Fradkin}}\ and\ \bibinfo {author} {\bibfnamefont {S.~A.}\ \bibnamefont
  {Kivelson}},\ }\href {\doibase 10.1103/PhysRevB.59.8065} {\bibfield
  {journal} {\bibinfo  {journal} {Phys. Rev. B}\ }\textbf {\bibinfo {volume}
  {59}},\ \bibinfo {pages} {8065} (\bibinfo {year} {1999})}\BibitemShut
  {NoStop}%
\bibitem [{\citenamefont {Barci}\ \emph {et~al.}(2002)\citenamefont {Barci},
  \citenamefont {Fradkin}, \citenamefont {Kivelson},\ and\ \citenamefont
  {Oganesyan}}]{BaFrKiOg2002}%
  \BibitemOpen
  \bibfield  {author} {\bibinfo {author} {\bibfnamefont {D.~G.}\ \bibnamefont
  {Barci}}, \bibinfo {author} {\bibfnamefont {E.}~\bibnamefont {Fradkin}},
  \bibinfo {author} {\bibfnamefont {S.~A.}\ \bibnamefont {Kivelson}}, \ and\
  \bibinfo {author} {\bibfnamefont {V.}~\bibnamefont {Oganesyan}},\ }\href@noop
  {} {\bibfield  {journal} {\bibinfo  {journal} {Phys. Rev. B}\ }\textbf
  {\bibinfo {volume} {65}},\ \bibinfo {pages} {245319} (\bibinfo {year}
  {2002})}\BibitemShut {NoStop}%
\bibitem [{\citenamefont {Lawler}\ \emph {et~al.}(2006)\citenamefont {Lawler},
  \citenamefont {Barci}, \citenamefont {Fern\'andez}, \citenamefont {Fradkin},\
  and\ \citenamefont {Oxman}}]{Lawler2006}%
  \BibitemOpen
  \bibfield  {author} {\bibinfo {author} {\bibfnamefont {M.~J.}\ \bibnamefont
  {Lawler}}, \bibinfo {author} {\bibfnamefont {D.~G.}\ \bibnamefont {Barci}},
  \bibinfo {author} {\bibfnamefont {V.}~\bibnamefont {Fern\'andez}}, \bibinfo
  {author} {\bibfnamefont {E.}~\bibnamefont {Fradkin}}, \ and\ \bibinfo
  {author} {\bibfnamefont {L.}~\bibnamefont {Oxman}},\ }\href {\doibase
  10.1103/PhysRevB.73.085101} {\bibfield  {journal} {\bibinfo  {journal} {Phys.
  Rev. B}\ }\textbf {\bibinfo {volume} {73}},\ \bibinfo {pages} {085101}
  (\bibinfo {year} {2006})}\BibitemShut {NoStop}%
\bibitem [{\citenamefont {Seul}\ \emph {et~al.}(1991)\citenamefont {Seul},
  \citenamefont {Monar}, \citenamefont {O'Gorman},\ and\ \citenamefont
  {Wolfe}}]{SeMoGoWo1991}%
  \BibitemOpen
  \bibfield  {author} {\bibinfo {author} {\bibfnamefont {M.}~\bibnamefont
  {Seul}}, \bibinfo {author} {\bibfnamefont {L.~R.}\ \bibnamefont {Monar}},
  \bibinfo {author} {\bibfnamefont {L.}~\bibnamefont {O'Gorman}}, \ and\
  \bibinfo {author} {\bibfnamefont {R.}~\bibnamefont {Wolfe}},\ }\href@noop {}
  {\bibfield  {journal} {\bibinfo  {journal} {Science}\ }\textbf {\bibinfo
  {volume} {254}},\ \bibinfo {pages} {1616} (\bibinfo {year}
  {1991})}\BibitemShut {NoStop}%
\bibitem [{\citenamefont {Vega}\ \emph {et~al.}(2005)\citenamefont {Vega},
  \citenamefont {Harrison}, \citenamefont {Angelescu}, \citenamefont {Trawick},
  \citenamefont {Huse}, \citenamefont {Chaikin},\ and\ \citenamefont
  {Register}}]{VeHaAnTrHuChRe2005}%
  \BibitemOpen
  \bibfield  {author} {\bibinfo {author} {\bibfnamefont {D.~A.}\ \bibnamefont
  {Vega}}, \bibinfo {author} {\bibfnamefont {C.~K.}\ \bibnamefont {Harrison}},
  \bibinfo {author} {\bibfnamefont {D.~E.}\ \bibnamefont {Angelescu}}, \bibinfo
  {author} {\bibfnamefont {M.~L.}\ \bibnamefont {Trawick}}, \bibinfo {author}
  {\bibfnamefont {D.~A.}\ \bibnamefont {Huse}}, \bibinfo {author}
  {\bibfnamefont {P.~M.}\ \bibnamefont {Chaikin}}, \ and\ \bibinfo {author}
  {\bibfnamefont {R.~A.}\ \bibnamefont {Register}},\ }\href {\doibase
  10.1103/PhysRevE.71.061803} {\bibfield  {journal} {\bibinfo  {journal} {Phys.
  Rev. E}\ }\textbf {\bibinfo {volume} {71}},\ \bibinfo {pages} {061803}
  (\bibinfo {year} {2005})}\BibitemShut {NoStop}%
\bibitem [{\citenamefont {Ruiz}\ \emph {et~al.}(2008)\citenamefont {Ruiz},
  \citenamefont {Bosworth},\ and\ \citenamefont {Black}}]{RuBoBl2008}%
  \BibitemOpen
  \bibfield  {author} {\bibinfo {author} {\bibfnamefont {R.}~\bibnamefont
  {Ruiz}}, \bibinfo {author} {\bibfnamefont {J.~K.}\ \bibnamefont {Bosworth}},
  \ and\ \bibinfo {author} {\bibfnamefont {C.~T.}\ \bibnamefont {Black}},\
  }\href {\doibase 10.1103/PhysRevB.77.054204} {\bibfield  {journal} {\bibinfo
  {journal} {Physical Review B}\ }\textbf {\bibinfo {volume} {77}},\ \bibinfo
  {eid} {054204} (\bibinfo {year} {2008})}\BibitemShut {NoStop}%
\bibitem [{\citenamefont {Malescio}\ and\ \citenamefont
  {Pellicane}(2004)}]{MaPe2004}%
  \BibitemOpen
  \bibfield  {author} {\bibinfo {author} {\bibfnamefont {G.}~\bibnamefont
  {Malescio}}\ and\ \bibinfo {author} {\bibfnamefont {G.}~\bibnamefont
  {Pellicane}},\ }\href {\doibase 10.1103/PhysRevE.70.021202} {\bibfield
  {journal} {\bibinfo  {journal} {Phys. Rev. E}\ }\textbf {\bibinfo {volume}
  {70}},\ \bibinfo {pages} {021202} (\bibinfo {year} {2004})}\BibitemShut
  {NoStop}%
\bibitem [{\citenamefont {Imperio}\ and\ \citenamefont
  {Reatto}(2006)}]{ImRe2006}%
  \BibitemOpen
  \bibfield  {author} {\bibinfo {author} {\bibfnamefont {A.}~\bibnamefont
  {Imperio}}\ and\ \bibinfo {author} {\bibfnamefont {L.}~\bibnamefont
  {Reatto}},\ }\href {\doibase 10.1063/1.2185618} {\bibfield  {journal}
  {\bibinfo  {journal} {The Journal of Chemical Physics}\ }\textbf {\bibinfo
  {volume} {124}},\ \bibinfo {eid} {164712} (\bibinfo {year}
  {2006})}\BibitemShut {NoStop}%
\bibitem [{\citenamefont {Glaser}\ \emph {et~al.}(2007)\citenamefont {Glaser},
  \citenamefont {Grason}, \citenamefont {Kamien}, \citenamefont {Ko{\v s}mrlj},
  \citenamefont {Santangelo},\ and\ \citenamefont {Ziherl}}]{GlGrKaKoSaZi2007}%
  \BibitemOpen
  \bibfield  {author} {\bibinfo {author} {\bibfnamefont {M.~A.}\ \bibnamefont
  {Glaser}}, \bibinfo {author} {\bibfnamefont {G.~M.}\ \bibnamefont {Grason}},
  \bibinfo {author} {\bibfnamefont {R.~D.}\ \bibnamefont {Kamien}}, \bibinfo
  {author} {\bibfnamefont {A.}~\bibnamefont {Ko{\v s}mrlj}}, \bibinfo {author}
  {\bibfnamefont {C.~D.}\ \bibnamefont {Santangelo}}, \ and\ \bibinfo {author}
  {\bibfnamefont {P.}~\bibnamefont {Ziherl}},\ }\href@noop {} {\bibfield
  {journal} {\bibinfo  {journal} {EPL (Europhysics Letters)}\ }\textbf
  {\bibinfo {volume} {78}},\ \bibinfo {pages} {46004} (\bibinfo {year}
  {2007})}\BibitemShut {NoStop}%
\bibitem [{\citenamefont {Seul}\ and\ \citenamefont
  {Andelman}(1995)}]{SeAn1995}%
  \BibitemOpen
  \bibfield  {author} {\bibinfo {author} {\bibfnamefont {M.}~\bibnamefont
  {Seul}}\ and\ \bibinfo {author} {\bibfnamefont {D.}~\bibnamefont
  {Andelman}},\ }\href@noop {} {\bibfield  {journal} {\bibinfo  {journal}
  {Science}\ }\textbf {\bibinfo {volume} {267}},\ \bibinfo {pages} {476}
  (\bibinfo {year} {1995})}\BibitemShut {NoStop}%
\bibitem [{\citenamefont {Barci}\ and\ \citenamefont
  {Stariolo}(2007)}]{BaSt2007}%
  \BibitemOpen
  \bibfield  {author} {\bibinfo {author} {\bibfnamefont {D.~G.}\ \bibnamefont
  {Barci}}\ and\ \bibinfo {author} {\bibfnamefont {D.~A.}\ \bibnamefont
  {Stariolo}},\ }\href {\doibase 10.1103/PhysRevLett.98.200604} {\bibfield
  {journal} {\bibinfo  {journal} {Physical Review Letters}\ }\textbf {\bibinfo
  {volume} {98}},\ \bibinfo {eid} {200604} (\bibinfo {year}
  {2007})}\BibitemShut {NoStop}%
\bibitem [{\citenamefont {Brazovskii}(1975)}]{Br1975}%
  \BibitemOpen
  \bibfield  {author} {\bibinfo {author} {\bibfnamefont {S.~A.}\ \bibnamefont
  {Brazovskii}},\ }\href@noop {} {\bibfield  {journal} {\bibinfo  {journal}
  {Sov. Phys. JETP}\ }\textbf {\bibinfo {volume} {41}},\ \bibinfo {pages} {85}
  (\bibinfo {year} {1975})}\BibitemShut {NoStop}%
\bibitem [{\citenamefont {Barci}\ and\ \citenamefont
  {Stariolo}(2009)}]{BaSt2009}%
  \BibitemOpen
  \bibfield  {author} {\bibinfo {author} {\bibfnamefont {D.~G.}\ \bibnamefont
  {Barci}}\ and\ \bibinfo {author} {\bibfnamefont {D.~A.}\ \bibnamefont
  {Stariolo}},\ }\href {\doibase 10.1103/PhysRevB.79.075437} {\bibfield
  {journal} {\bibinfo  {journal} {Physical Review B}\ }\textbf {\bibinfo
  {volume} {79}},\ \bibinfo {eid} {075437} (\bibinfo {year}
  {2009})}\BibitemShut {NoStop}%
\bibitem [{\citenamefont {Barci}\ and\ \citenamefont
  {Stariolo}(2011)}]{BaSt2011}%
  \BibitemOpen
  \bibfield  {author} {\bibinfo {author} {\bibfnamefont {D.~G.}\ \bibnamefont
  {Barci}}\ and\ \bibinfo {author} {\bibfnamefont {D.~A.}\ \bibnamefont
  {Stariolo}},\ }\href {\doibase 10.1103/PhysRevB.84.094439} {\bibfield
  {journal} {\bibinfo  {journal} {Phys. Rev. B}\ }\textbf {\bibinfo {volume}
  {84}},\ \bibinfo {pages} {094439} (\bibinfo {year} {2011})}\BibitemShut
  {NoStop}%
\bibitem [{\citenamefont {Cannas}\ \emph {et~al.}(2006)\citenamefont {Cannas},
  \citenamefont {Michelon}, \citenamefont {Stariolo},\ and\ \citenamefont
  {Tamarit}}]{CaMiStTa2006}%
  \BibitemOpen
  \bibfield  {author} {\bibinfo {author} {\bibfnamefont {S.~A.}\ \bibnamefont
  {Cannas}}, \bibinfo {author} {\bibfnamefont {M.~F.}\ \bibnamefont
  {Michelon}}, \bibinfo {author} {\bibfnamefont {D.~A.}\ \bibnamefont
  {Stariolo}}, \ and\ \bibinfo {author} {\bibfnamefont {F.~A.}\ \bibnamefont
  {Tamarit}},\ }\href {\doibase 10.1103/PhysRevB.73.184425} {\bibfield
  {journal} {\bibinfo  {journal} {Phys. Rev. B}\ }\textbf {\bibinfo {volume}
  {73}},\ \bibinfo {pages} {184425} (\bibinfo {year} {2006})}\BibitemShut
  {NoStop}%
\bibitem [{\citenamefont {Jin}\ \emph {et~al.}(2012)\citenamefont {Jin},
  \citenamefont {Sen},\ and\ \citenamefont {Sandvik}}]{JiSeSa2012}%
  \BibitemOpen
  \bibfield  {author} {\bibinfo {author} {\bibfnamefont {S.}~\bibnamefont
  {Jin}}, \bibinfo {author} {\bibfnamefont {A.}~\bibnamefont {Sen}}, \ and\
  \bibinfo {author} {\bibfnamefont {A.~W.}\ \bibnamefont {Sandvik}},\ }\href
  {\doibase 10.1103/PhysRevLett.108.045702} {\bibfield  {journal} {\bibinfo
  {journal} {Phys. Rev. Lett.}\ }\textbf {\bibinfo {volume} {108}},\ \bibinfo
  {pages} {045702} (\bibinfo {year} {2012})}\BibitemShut {NoStop}%
\bibitem [{\citenamefont {Grousson}\ \emph {et~al.}(2000)\citenamefont
  {Grousson}, \citenamefont {Tarjus},\ and\ \citenamefont {Viot}}]{GrTaVi2000}%
  \BibitemOpen
  \bibfield  {author} {\bibinfo {author} {\bibfnamefont {M.}~\bibnamefont
  {Grousson}}, \bibinfo {author} {\bibfnamefont {G.}~\bibnamefont {Tarjus}}, \
  and\ \bibinfo {author} {\bibfnamefont {P.}~\bibnamefont {Viot}},\ }\href
  {\doibase 10.1103/PhysRevE.62.7781} {\bibfield  {journal} {\bibinfo
  {journal} {Phys. Rev. E}\ }\textbf {\bibinfo {volume} {62}},\ \bibinfo
  {pages} {7781} (\bibinfo {year} {2000})}\BibitemShut {NoStop}%
\bibitem [{\citenamefont {Barr\'e}\ \emph {et~al.}(2001)\citenamefont
  {Barr\'e}, \citenamefont {Mukamel},\ and\ \citenamefont {Ruffo}}]{Ruffo2001}%
  \BibitemOpen
  \bibfield  {author} {\bibinfo {author} {\bibfnamefont {J.}~\bibnamefont
  {Barr\'e}}, \bibinfo {author} {\bibfnamefont {D.}~\bibnamefont {Mukamel}}, \
  and\ \bibinfo {author} {\bibfnamefont {S.}~\bibnamefont {Ruffo}},\ }\href
  {\doibase 10.1103/PhysRevLett.87.030601} {\bibfield  {journal} {\bibinfo
  {journal} {Phys. Rev. Lett.}\ }\textbf {\bibinfo {volume} {87}},\ \bibinfo
  {pages} {030601} (\bibinfo {year} {2001})}\BibitemShut {NoStop}%
\bibitem [{\citenamefont {Barr\'e}\ \emph {et~al.}(2005)\citenamefont
  {Barr\'e}, \citenamefont {Bouchet}, \citenamefont {Dauxois},\ and\
  \citenamefont {Ruffo}}]{Ruffo2005}%
  \BibitemOpen
  \bibfield  {author} {\bibinfo {author} {\bibfnamefont {J.}~\bibnamefont
  {Barr\'e}}, \bibinfo {author} {\bibfnamefont {F.}~\bibnamefont {Bouchet}},
  \bibinfo {author} {\bibfnamefont {T.}~\bibnamefont {Dauxois}}, \ and\
  \bibinfo {author} {\bibfnamefont {S.}~\bibnamefont {Ruffo}},\ }\href@noop {}
  {\bibfield  {journal} {\bibinfo  {journal} {Journal of Statistical Physics}\
  }\textbf {\bibinfo {volume} {119}},\ \bibinfo {pages} {677} (\bibinfo {year}
  {2005})}\BibitemShut {NoStop}%
\bibitem [{\citenamefont {Binney}\ \emph {et~al.}(1995)\citenamefont {Binney},
  \citenamefont {Dowrick}, \citenamefont {Fisher},\ and\ \citenamefont
  {Newman}}]{BiDoFiNe1995}%
  \BibitemOpen
  \bibfield  {author} {\bibinfo {author} {\bibfnamefont {J.~J.}\ \bibnamefont
  {Binney}}, \bibinfo {author} {\bibfnamefont {N.~J.}\ \bibnamefont {Dowrick}},
  \bibinfo {author} {\bibfnamefont {A.~J.}\ \bibnamefont {Fisher}}, \ and\
  \bibinfo {author} {\bibfnamefont {M.~E.~J.}\ \bibnamefont {Newman}},\
  }\href@noop {} {\emph {\bibinfo {title} {The Theory of Critical Phenomena}}}\
  (\bibinfo  {publisher} {Oxford University Press},\ \bibinfo {year}
  {1995})\BibitemShut {NoStop}%
\bibitem [{\citenamefont {Amit}(1978)}]{Amit1978}%
  \BibitemOpen
  \bibfield  {author} {\bibinfo {author} {\bibfnamefont {D.~J.}\ \bibnamefont
  {Amit}},\ }\href@noop {} {\emph {\bibinfo {title} {{Field Theory, the
  Renormalization Group, and Critical Phenomena}}}}\ (\bibinfo  {publisher}
  {McGraw-Hill International Book Company},\ \bibinfo {address} {New York},\
  \bibinfo {year} {1978})\BibitemShut {NoStop}%
\bibitem [{\citenamefont {Gel'fand}\ and\ \citenamefont
  {Shilov}(1964)}]{Gelfand}%
  \BibitemOpen
  \bibfield  {author} {\bibinfo {author} {\bibfnamefont {I.~M.}\ \bibnamefont
  {Gel'fand}}\ and\ \bibinfo {author} {\bibfnamefont {G.~E.}\ \bibnamefont
  {Shilov}},\ }\href@noop {} {\emph {\bibinfo {title} {Generalized
  Functions}}},\ Vol.~\bibinfo {volume} {I}\ (\bibinfo  {publisher} {Academic
  Press},\ \bibinfo {year} {1964})\BibitemShut {NoStop}%
\bibitem [{\citenamefont {Portmann}\ \emph {et~al.}(2010)\citenamefont
  {Portmann}, \citenamefont {G\"olzer}, \citenamefont {Saratz}, \citenamefont
  {Billoni}, \citenamefont {Pescia},\ and\ \citenamefont
  {Vindigni}}]{Portman2010}%
  \BibitemOpen
  \bibfield  {author} {\bibinfo {author} {\bibfnamefont {O.}~\bibnamefont
  {Portmann}}, \bibinfo {author} {\bibfnamefont {A.}~\bibnamefont {G\"olzer}},
  \bibinfo {author} {\bibfnamefont {N.}~\bibnamefont {Saratz}}, \bibinfo
  {author} {\bibfnamefont {O.~V.}\ \bibnamefont {Billoni}}, \bibinfo {author}
  {\bibfnamefont {D.}~\bibnamefont {Pescia}}, \ and\ \bibinfo {author}
  {\bibfnamefont {A.}~\bibnamefont {Vindigni}},\ }\href {\doibase
  10.1103/PhysRevB.82.184409} {\bibfield  {journal} {\bibinfo  {journal} {Phys.
  Rev. B}\ }\textbf {\bibinfo {volume} {82}},\ \bibinfo {pages} {184409}
  (\bibinfo {year} {2010})}\BibitemShut {NoStop}%
\end{thebibliography}

\end{document}